\documentclass[10pt,letterpaper]{article}
\usepackage[top=0.85in,left=2.75in,footskip=0.75in]{geometry}

\usepackage{amsmath,amssymb}

\usepackage{changepage}

\usepackage[utf8x]{inputenc}

\usepackage{textcomp,marvosym}

\usepackage{cite}

\usepackage{nameref,hyperref}

\usepackage[right]{lineno}

\usepackage{microtype}
\DisableLigatures[f]{encoding = *, family = * }

\usepackage[table]{xcolor}

\usepackage{array}

\newcolumntype{+}{!{\vrule width 2pt}}

\newlength\savedwidth

\raggedright
\usepackage[aboveskip=1pt,labelfont=bf,labelsep=period,justification=raggedright,singlelinecheck=off]{caption}

\makeatletter
\renewcommand{\@biblabel}[1]{\quad#1.}
\makeatother

\usepackage{lastpage,fancyhdr,graphicx}
\usepackage{epstopdf}
\fancyheadoffset[L]{2.25in}
\fancyfootoffset[L]{2.25in}
\begin{document}
\vspace*{0.2in}

\begin{flushleft}
{\Large
\textbf\newline{Design principles for the glycoprotein quality control pathway} 
}
\newline
\\
Aidan I. Brown\textsuperscript{1},
Elena F. Koslover\textsuperscript{1*},
\\
\bigskip
\textbf{1} Department of Physics, University of California, San Diego, San Diego, California 92093
\bigskip

%
%





* ekoslover@ucsd.edu

\end{flushleft}
\section*{Abstract}
Newly-translated glycoproteins in the endoplasmic reticulum (ER) often undergo cycles of chaperone binding and release in order to assist in folding. Quality control is required to distinguish between proteins that have completed native folding, those that have yet to fold, and those that have misfolded. 
Using quantitative modeling, we explore how the design of the quality-control pathway modulates its efficiency.
Our results show that an energy-consuming cyclic quality-control process, 
similar to the observed physiological system,
outperforms alternative designs. The kinetic parameters that optimize the performance of this system drastically change with protein production levels, while remaining relatively insensitive to the protein folding rate. Adjusting only the degradation rate, while fixing other parameters, allows the pathway to adapt across a range of protein production levels, aligning with
\textit{in vivo}
measurements that implicate the release of degradation-associated enzymes as a rapid-response system for perturbations in protein  homeostasis. The quantitative models developed here elucidate design principles for effective glycoprotein quality control in the ER, improving our mechanistic understanding of a system crucial to maintaining cellular health.

\section*{Author summary}

We explore the architecture and limitations of the quality-control pathway responsible for efficient folding of secretory proteins. Newly-synthesized proteins are tagged by the attachment of a `glycan' sugar chain which facilitates their binding to a chaperone that assists protein folding. Removal of a specific sugar group on the glycan triggers release from the chaperone, and not-yet-folded proteins can be re-tagged for another round of chaperone binding.
A degradation pathway acts in parallel with the folding cycle, to remove those proteins that have remained unfolded for a sufficiently long time.
We develop and solve a mathematical model of this quality-control system, showing that the cyclical design found in living cells is uniquely able to 
maximize folded protein throughput while avoiding accumulation of unfolded proteins.
Although this physiological model provides the best performance, its parameters must be adjusted to perform optimally under different protein production loads, and any single fixed set of parameters leads to poor performance when production rate is altered. We find that a single adjustable parameter, the protein degradation rate, is sufficient to allow optimal performance across a range of conditions. Interestingly, observations of living cells suggest that the degradation speed is indeed rapidly adjusted.


\section*{Introduction}

The general principle of quality control is of critical importance to the maintenance, function, and growth of biological cells. Autophagy and the ubiquitin-proteasome system selectively remove damaged proteins and organelles to maintain the quality of cellular components~\cite{murrow2013autophagy,pohl2019cellular}. Fidelity is aided by proofreading processes during DNA copying~\cite{hopfield1974kinetic}, immune  signaling~\cite{mckeithan1995kinetic}, and external sensing~\cite{hartich2015nonequilibrium}. Quality control is particularly important for proteins, with a high fraction of proteome mass across the kingdoms of life devoted to protein homeostasis and folding~\cite{muller2020}. Unfolded and misfolded proteins often form aggregates, which can impede cellular processes and are associated with a variety of human diseases~\cite{mogk2018cellular,dobson1999protein,soto2003unfolding,gregersen2006protein}.

Protein quality control begins with transcriptional proofreading by RNA polymerase~\cite{mellenius2017transcriptional} and continues with proofreading of tRNA matching to mRNA codons during translation~\cite{sharma2011distribution} to reduce errors in the polypeptide sequence. Quality control continues beyond production, throughout the lifetime of a protein~\cite{rodrigo2012design,wolff2014differential,shao2016target}. We focus on  post-translational quality control pathways that ensure nascent polypeptides fold into the correct or `native' three-dimensional conformation, rather than roaming the cell in a misfolded state~\cite{rodrigo2012design,wolff2014differential,shao2016target}. 

Nearly one-third of eukaryotic proteins, or $\sim$8000 proteins in humans, are synthesized through the secretory pathway and begin as nascent polypetides in the endoplasmic reticulum (ER)~\cite{caramelo2015sweet}. The majority of ER-manufactured proteins acquire branched carbohydrate chains, via N-linked glycosylation~\cite{adams2019protein}.
While these glycan chains can be important for protein function~\cite{dwek1996glycobiology} and stabilization~\cite{shental2008effect}, the specific sugar residues in the glycan serve as a tunable barcode to direct the interactions that lead to further protein folding attempts or protein degradation~\cite{caramelo2015sweet}. Accordingly, glycans play a key role in the folding quality control of secretory proteins.

The quality control pathway, in deciding which proteins to degrade and which to continue folding, attempts to distinguish between three groups of proteins: natively folded, as yet unfolded, and terminally misfolded. Natively folded proteins can be distinguished by the lack of exposed hydrophobic residues and free thiols~\cite{ellgaard1999setting,anelli2008protein}, and are permitted to leave the ER to continue through the secretory pathway. It is less straightforward to distinguish between as yet unfolded proteins, which should be provided more time to fold; and terminally misfolded proteins, which should be targeted for degradation~\cite{tannous2015reglucosylation}. 
Newly-synthesized proteins and unfolded proteins are flagged by monoglucosylation of a glycan chain, which facilitates chaperone binding to attempt folding. Proteins dissociate from the chaperone upon removal of this glucose moiety, which is not added back to proteins that have reached their native conformation. Proteins that fail to reach a native conformation will eventually experience trimming of other glycan moieties, leading to degradation via the ER-associated degradation (ERAD) pathway~\cite{caramelo2015sweet}.

In this work we investigate how the design of the glycoprotein folding quality-control pathway facilitates decisions of whether nascent proteins may continue trying to fold, and how specific pathway features impact performance. Specifically, we seek to understand the advantages provided by the cyclic structure of the quality control pathway. Overall, we find that the consensus physiological model outperforms other designs, and describe how its kinetic parameters  can be tuned to maintain performance across a broad range of conditions.

\section*{Model}

Upon translation, glycoproteins enter the quality control pathway marked with a single glucose moiety~\cite{aebi2010n,caramelo2015sweet}. These monoglucosylated proteins can bind calnexin and calreticulin~\cite{molinari2007n}, chaperone proteins that assist protein folding. Glycoproteins are released from chaperones upon trimming of the glucose by glucosidase II~\cite{hebert1995glucose,tannous2015reglucosylation,lamriben2016n,shenkman2019compartmentalization}.

Proteins that have reached a native conformation are eligible to be exported from the ER and to proceed down the secretory pathway~\cite{budnik2009er}. However, not all proteins that are released from the chaperone are successfully folded.
Uridine diphosphate-glucose:glycoprotein glucosyltransferase (UGGT) can reglucosylate incompletely folded glycoproteins to enable another round of chaperone binding that further facilitates folding~\cite{lamriben2016n,kiuchi2018monitoring}.
UGGT does not reglucosylate proteins that have reached a native conformation, and is thought to use indicators such as the availability of the entire glycan chain and hydrophobic patches to detect non-native conformations~\cite{caramelo2015sweet,tannous2015n,lamriben2016n,kiuchi2018monitoring}. There is some evidence that UGGT may prefer to reglucosylate unfolded glycoproteins rather than those that have misfolded into an incorrect conformation, but overall it is unclear if UGGT can distinguish between these two groups of non-natively folded proteins~\cite{tannous2015reglucosylation,tannous2015n,kiuchi2018monitoring}.

Glycoprotein interaction with chaperones, glucosidase II, and UGGT thus forms a cycle: a monoglucosylated protein binds a chaperone (calnexin or calreticulin) for folding assistance, the glucose is trimmed by glucosidase II to release the protein from the chaperone, and UGGT restores the glucose to non-natively folded proteins to direct chaperone rebinding~\cite{kiuchi2018monitoring}. Folding time in the ER can vary from a few minutes to several hours~\cite{hebert2007and}, with some proteins natively folded after one round of chaperone binding, and others requiring multiple rounds of chaperone interaction~\cite{lamriben2016n}.

In addition to departing the cycle by folding, proteins can be selected for ER-associated degradation (ERAD), a pathway involving removal from the ER followed by proteasomal degradation~\cite{lamriben2016n}. 
Commitment of a protein to the ERAD pathway for degradation can involve interaction with various enzymes, some of which irreversibly trim additional moieties off the glycan chains~\cite{brambilla2016five,spiro1996definition,caramelo2015sweet,sousa1992recognition,molinari2007n,quan2008defining,christianson20089,hosokawa2009human,clerc2009htm1,avezov2008endoplasmic,tannous2015reglucosylation,shenkman2018mannosidase,adams2019protein}. 
Unglucosylated glycans, which do not allow chaperone binding, are thought to be specifically vulnerable to the modifications that commit a protein to ERAD~\cite{frenkel2003endoplasmic,ellgaard2003quality,avezov2008endoplasmic}.

\begin{figure}
	\begin{center}
	\includegraphics[width=0.67\textwidth]{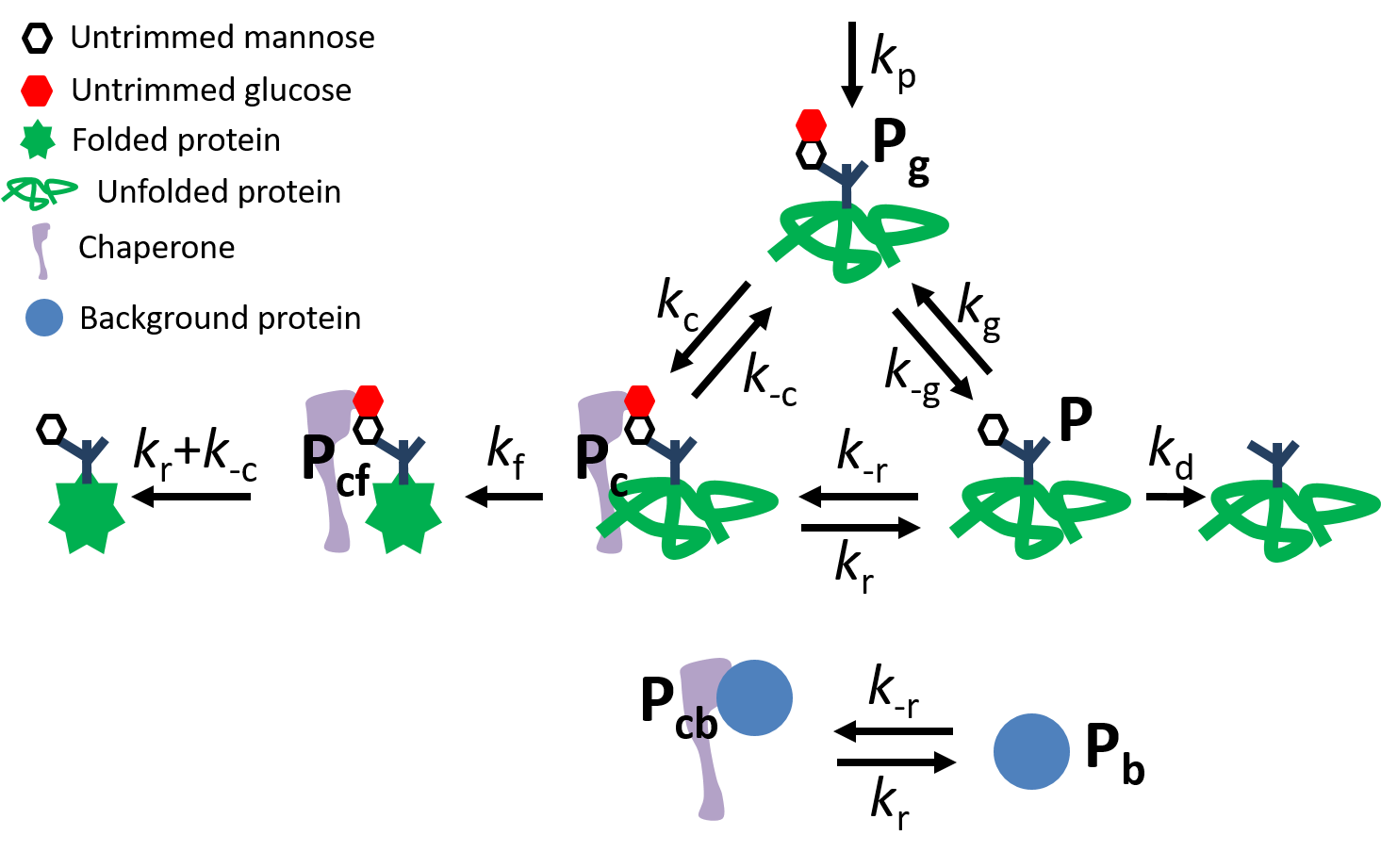}
	\end{center}
	\caption{{\bf Model of glycoprotein quality control via the chaperone binding cycle.} $P_g$ represents the monoglucosylated proteins, $P_{\text{c}}$ the unfolded chaperone-bound proteins, $P_{\text{cf}}$ the folded chaperone-bound proteins, $P$ the proteins lacking a glucose tag, $P_{\text{b}}$ the background proteins, and $P_{\text{cb}}$ the chaperone-bound background proteins.
	}
	\label{fig:cycle}
\end{figure}

We represent the glycoprotein quality control cycle with three discrete states, along with an additional discrete state for chaperone-bound natively folded proteins (see Fig.~\ref{fig:cycle}). Proteins enter the cycle in a monoglucosylated state (whose concentration is represented by $P_{\text{g}}$) with a production rate $k_{\text{p}}$. Monoglucosylated proteins bind to chaperones as a bimolecular reaction with rate constant $k_{\text{c}}$. The available chaperone concentration is represented by $C_{\text{A}}$  and the concentration of chaperone-bound unfolded proteins by $P_{\text{c}}$. Proteins bound to the chaperone fold into their native conformation with rate constant $k_{\text{f}}$, and $P_{\text{cf}}$ represents the concentration of folded proteins bound to the chaperone. Chaperone-bound proteins (both natively folded and not) are removed from the chaperone with rate constant $k_{\text{r}}$, with natively folded proteins then exiting the cycle. Proteins removed from the chaperone that are not natively folded (at concentration $P$) are lacking a glucose moiety, and can be reglucosylated with a rate constant $k_{\text{g}}$. Monoglucosylated proteins not bound to a chaperone can have their glucose removed with rate constant $k_{\text{-g}}$, serving as a ``safety-valve" pathway when the concentration of proteins to be folded overwhelms the available chaperones. 
Deglucosylated proteins are vulnerable to degradation via ERAD~\cite{frenkel2003endoplasmic,ellgaard2003quality,avezov2008endoplasmic}. Specifically, the sugar moiety to which the glucose attaches can be removed, irreversibly committing the protein to degradation via the ERAD pathway~\cite{ermonval2001n,olivari2006edem1,hebert2007and,caramelo2008getting,avezov2008endoplasmic,caramelo2015sweet,shenkman2018mannosidase,adams2019protein}. We treat ERAD commitment and protein degradation as a single irreversible process with rate constant $k_{\text{d}}$.

Although not discussed in the consensus physiological model, for completeness we also consider unbinding of monoglucosylated proteins from the chaperone (rate constant $k_{\text{-c}}$) and rebinding of deglucosylated proteins back to the chaperone (rate constant $k_{\text{-r}}$).
Because such a putative rebinding pathway does not rely on a glucosylation signal to recognize proteins in need of folding, it is assumed to be non-specific and to allow the general binding of `background' proteins onto the chaperones. Such background proteins could include ER-resident proteins, or folded proteins that have not yet been exported.
The concentration of these additional background proteins
is represented by $P_\text{b}$ (for free background proteins) and $P_{\text{cb}}$ for background proteins bound to the chaperone. We assume each chaperone can bind only one protein at a time.

Overall, the dynamics of the chaperone binding cycle are described by
\begin{subequations}
	\label{eq:des}
	\begin{align}
		\frac{\text{d}P_{\text{g}}}{\text{d}t} &= k_{\text{g}}P + k_{\text{-c}}P_{\text{c}} - (k_{\text{c}}C_{\text{A}} + k_{\text{-g}})P_{\text{g}} + k_{\text{p}} \ , \\
		\frac{\text{d}P_{\text{c}}}{\text{d}t} &= (k_{\text{c}}P_{\text{g}} + k_{\text{-r}}P)C_{\text{A}} - (k_{\text{r}} + k_{\text{-c}} + k_{\text{f}})P_{\text{c}} \ , \\
		\frac{\text{d}P}{\text{d}t} &= k_{\text{r}}P_{\text{c}} + k_{\text{-g}}P_{\text{g}} - (k_{\text{g}} + k_{\text{-r}}C_{\text{A}} - k_{\text{d}})P \ , \\
		\frac{\text{d}P_{\text{cf}}}{\text{d}t} &= k_{\text{f}}P_{\text{c}} - (k_{\text{r}} + k_{\text{-c}})P_{\text{cf}} \ , \\
		\frac{\text{d}P_{\text{cb}}}{\text{d}t} &= k_{\text{-r}}C_{\text{A}}P_{\text{b}} - k_{\text{r}}P_{\text{cb}} \label{eq:delast} \ . 		
	\end{align}
\end{subequations}

Some proteins entering the chaperone binding cycle are unable to natively fold, as a result of translation errors or mutations~\cite{tyedmers2010cellular}. Heat and oxidative stress can also cause proteins to enter states that cannot fold~\cite{tyedmers2010cellular}, and these stressors may have a differential impact on different proteins.
We label these terminally misfolded, unfoldable proteins as simply `misfolded'. 
Their dynamics are described by
equations similar to Eq.~\ref{eq:des}a-c, with analogous protein quantities $P_{\text{g}}^*$, $P_{\text{c}}^*$, and $P^*$. 
The misfolded protein production rate is defined as $k_{\text{p}}^*$ and the folding rate is set to zero ($k_{\text{f}}^*=0$). All other rate constants are assumed to be identical for foldable and misfolded proteins.

Both the background proteins ($P_{\text{b}}$) and misfolded proteins ($P_i^*$) represent proteins capable of binding to and occupying the limited supply of total chaperone ($C_\text{tot}$) available in the cell. The concentration of available chaperones is then given by $C_{\text{A}} = C_{\text{tot}} - P_{\text{c}} - P_{\text{cf}} - P_{\text{c}}^* - P_{\text{cb}}$. In our model, background proteins represent those proteins that can bind weakly to the chaperone in the absence of a glucose moiety flagging them as newly-made proteins requiring folding. These can represent, for example, already folded proteins. They are not subject to the glucosylation and deglucosylation processes of the quality-control cycle. By contrast, `misfolded' proteins represent those that move through the quality control cycle with the same rate constants as normal proteins but are ultimately incapable of folding. In other words, the enzymes of the quality control cycle cannot distinguish these unfoldable proteins from native proteins~\cite{tannous2015reglucosylation,tannous2015n,kiuchi2018monitoring}.

The total rate of proteins entering the cycle is defined as $k_{\text{pt}} = k_{\text{p}} + k_{\text{p}}^*$ with a misfolded fraction $m_{\text{f}} = k_{\text{p}}^*/(k_{\text{p}} + k_{\text{p}}^*)$ unable to fold. Equations~\ref{eq:des} and the corresponding misfolded protein equations are non-dimensionalized by the timescale of glucose trimming for chaperone-bound proteins, setting $k_{\text{r}}=1$, and by total chaperone number, setting $C_{\text{tot}} = 1$ (see Methods for details).

For a given set of $k_i$, the steady state protein concentrations $P_i$ can be found, as derived in the Methods. We will use this steady-state solution to evaluate performance, on the assumption that protein production and processing parameters remain constant over timescales much longer than the individual cycle time.

\section*{Results}

\subsection*{Quality control efficiency and energy input} 

\begin{figure}
	\begin{center}
	\includegraphics[width=0.48\textwidth]{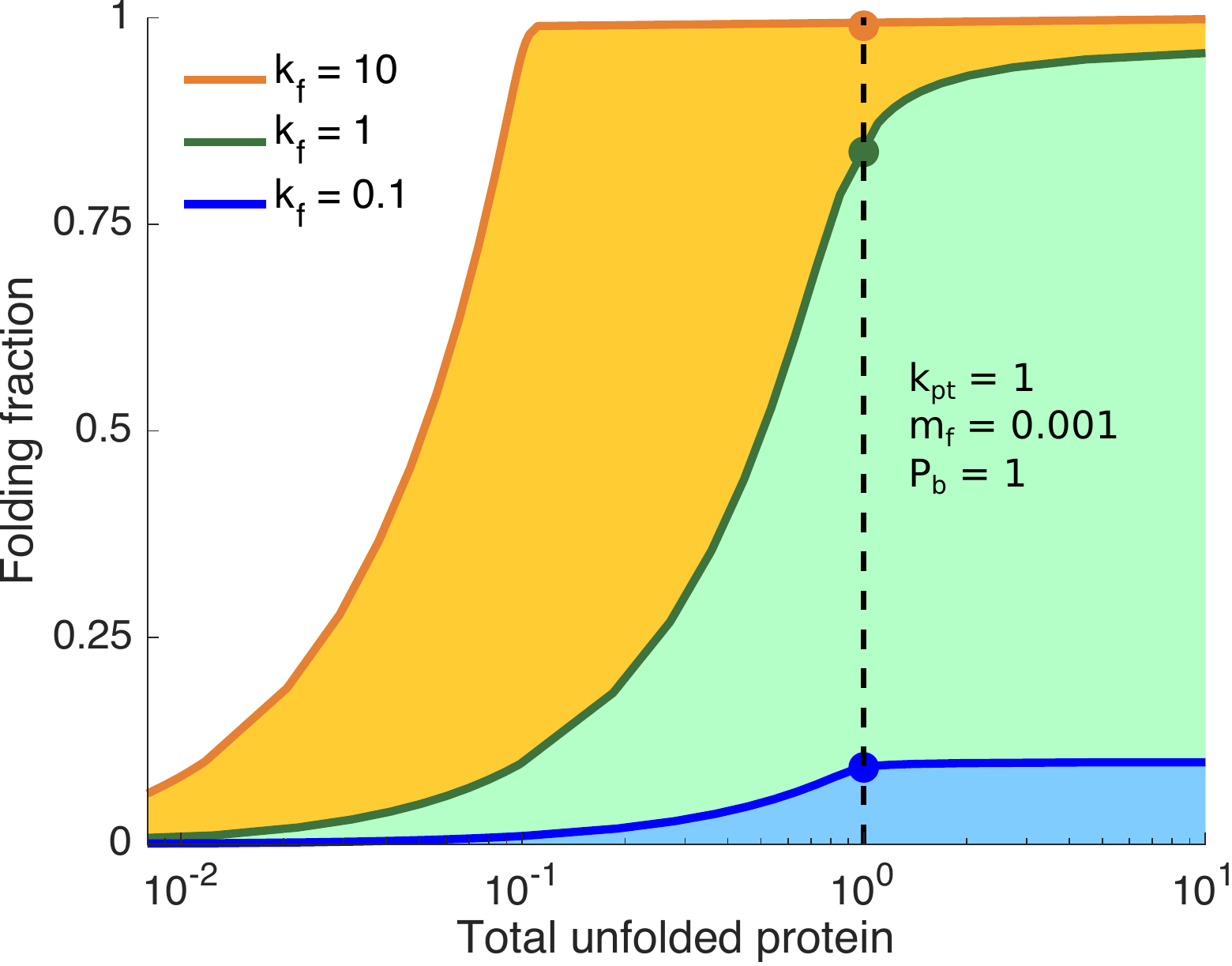}
	\end{center}
	\caption{
		{\bf Quality control engenders a trade-off between folding accuracy and speed.} Shaded regions in the phase diagram represent all combinations of folding fraction $f$ and total steady-state unfolded protein $P_\text{tot}$ that can be achieved by varying cycle parameters $k_{\text{c}}$, $k_{\text{-c}}$, $k_{\text{r}}$, $k_{\text{-r}}$, $k_{\text{g}}$, and $k_{\text{-g}}$, while keeping a fixed folding rate $k_\text{f}$, production rate $k_\text{pt}$, misfolded fraction $m_\text{f}$, and background protein concentration $P_{\text{b}}$. Solid lines represent the maximal achievable folding fraction $f_\text{max}$. Dots represent the efficiency metric $f_\text{max}^*$.	}
	\label{fig:phasediagram}
\end{figure}

We begin by considering how the glycoprotein folding system illustrated in Fig.~\ref{fig:cycle} is governed by a trade-off between accuracy and speed. On the one hand, the system needs to achieve robust, error-free quality control. On the other hand, it needs to process incoming proteins sufficiently rapidly to keep up with production and avoid accumulation of unfolded proteins in the cell.
We quantify system accuracy using the steady-state fraction of foldable proteins that successfully undergo folding rather than degradation,
\begin{equation}
	\label{eq:foldfrac}
	f = \frac{k_{\text{f}}P_{\text{c}}}{k_{\text{p}}} \ .
\end{equation}
A higher folding fraction $f$ indicates a more efficient folding process that produces more functional proteins per input of nascent unfolded proteins. 

A second metric for processing efficiency is the total unfolded protein present in the cycle at steady-state: $P_{\text{tot}} = P_{\text{g}} + P_{\text{g}}^* + P_{\text{c}} + P_{\text{c}}^* + P + P^*$. Low values of $P_\text{tot}$ correspond to rapid processing of individual nascent proteins that prevents their accumulation in the system.
High concentrations of unfolded proteins can lead to protein aggregation, which impede cellular function and health~\cite{mogk2018cellular}. With a typical influx to the ER of 0.1--1 million proteins per minute in each cell~\cite{karagoz2019unfolded}, proteins accumulate rapidly if the folding system cannot keep up with production.
High protein concentrations also induce ERAD and the unfolded protein response to limit the accumulation of protein aggregates, curtailing the throughput of functional proteins~\cite{hwang2018quality}.

Overall, we aim to understand how glycoprotein quality control can achieve both efficient shunting of foldable proteins towards folding rather than degradation, and rapid processing that limits the accumulation of unfolded proteins.
To assess this interplay, we determine the maximum folding fraction for each fixed value of total unfolded proteins, generating a phase-diagram of achievable values for these two metrics (Fig.~\ref{fig:phasediagram}). For fixed values of the production rate $k_\text{pt}$, misfolded fraction $m_\text{f}$,  folding rate $k_\text{f}$, and background protein level $P_{\text{b}}$, the cycle rate constants $k_{\text{c}}$, $k_{\text{-c}}$, $k_{\text{r}}$, $k_{\text{-r}}$, $k_{\text{g}}$, and $k_{\text{-g}}$ are allowed to vary (details in Methods) to map out the space of accessible efficiency metrics.
The curves of maximum folding fraction vs.\ total unfolded protein represent a Pareto frontier~\cite{shoval2012evolutionary} of folding cycle performance, where performance above or to the left of the curves in Fig.~\ref{fig:phasediagram} is not achievable. In Fig.~\ref{fig:phasediagram}, protein production ($k_{\text{pt}}$), misfolded fraction ($m_{\text{f}}$), and background protein concentration ($P_{\text{b}}$) are fixed for all curves, and each curve has a different protein folding speed ($k_{\text{f}}$).
Faster folding speeds allow for more efficient folding at each given value for the total unfolded protein.
The Pareto frontier has a characteristic shape of an increasing $f_{\text{max}}$ at low $P_{\text{tot}}$, followed by a plateau in $f_{\text{max}}$ at high $P_{\text{tot}}$. These curves demonstrate the trade-off between the two measures for efficient quality control, showing that maximization of folding fraction and minimization of total unfolded protein cannot be simultaneously achieved.

The characteristic curve shape in Fig.~\ref{fig:phasediagram} for $f_{\text{max}}$ vs.\ $P_{\text{tot}}$ suggests it is not always feasible to operate the glycoprotein quality control pathway at or near the maximum folding fraction as these high folding fractions can require a very high concentration of unfolded proteins. To assess pathway performance, we choose to limit the total unfolded protein quantity to $P_{\text{tot}}=1$, corresponding to a total unfolded protein concentration equal to the concentration of chaperones.
We then define the folding efficiency ($f_{\text{max}}^*$) as the maximum folding fraction at $P_{\text{tot}}=1$, serving as an overall utility function to evaluate the performance of the glycoprotein quality control pathway. 
This metric represents the best efficiency that can be achieved by the pathway without accumulating so many unfolded proteins as to overwhelm the binding capacity of the chaperones.

\begin{figure*}
	\includegraphics[width=0.96\textwidth]{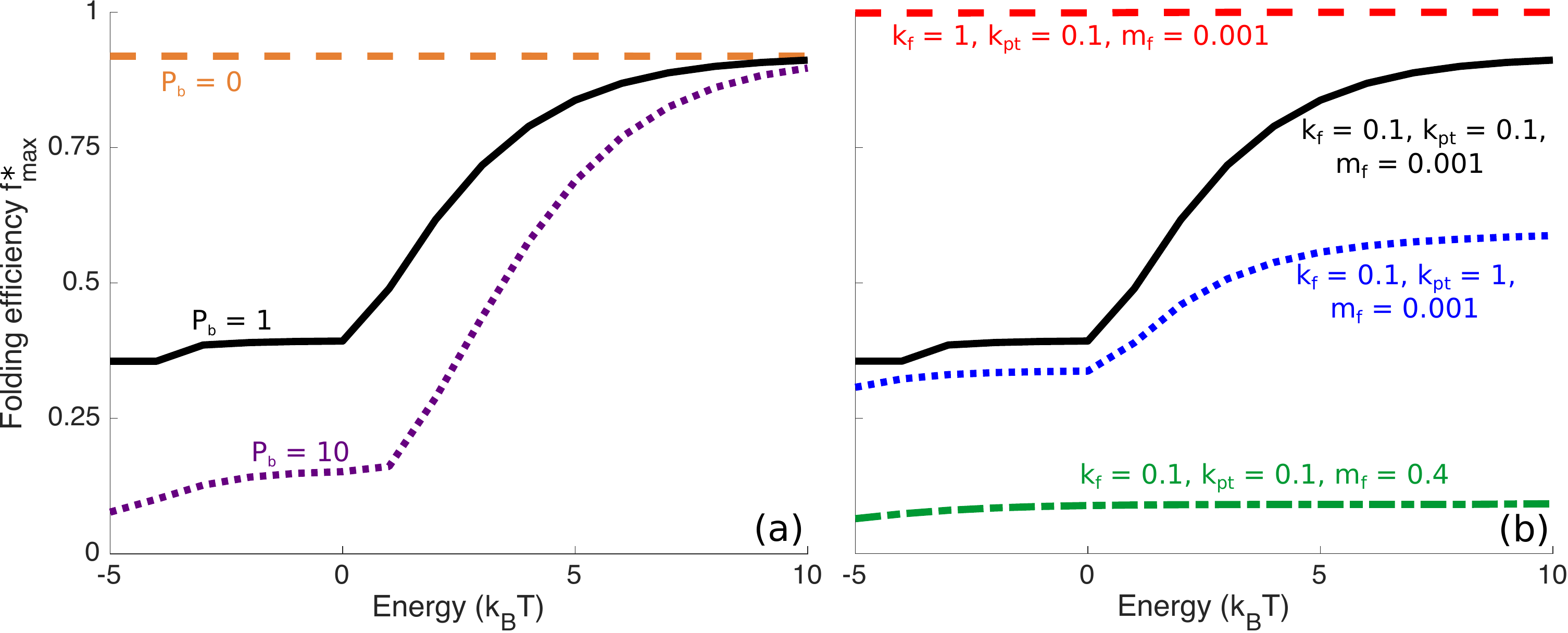}
	\caption{ 
		{\bf Nonequilibrium driving improves performance.}
		Each curve adjusts $k_{\text{c}}$, $k_{\text{-c}}$, $k_{\text{r}}$, $k_{\text{-r}}$, $k_{\text{g}}$, $k_{\text{-g}}$ to maximize the folding fraction (Eq.~\ref{eq:foldfrac}) while the total cycle energy (Eq.~\ref{eq:energy}) is varied and the total unfolded protein $P_{\text{tot}} = P_{\text{g}} + P_{\text{g}}^* + P_{\text{c}} + P_{\text{c}}^* + P + P^*$ is constrained to equal one. Other parameters are fixed for each curve.
		(a) Each curve shows a distinct level of background proteins $P_{\text{b}}$, with fixed $k_{\text{pt}}=0.1$, $k_{\text{f}}=0.1$, and $m_{\text{f}}=0.001$ for all curves.
		(b)
		Effects of increasing protein folding speed (red dashed), increasing protein production (blue dotted), and increasing misfolded fraction (green dashed-dotted) relative to the black curve, which is identical to the corresponding curve in (a). Background proteins are fixed to $P_{\text{b}}=1$ for all curves. 		
		%
	}
	\label{fig:energycycle}
\end{figure*}

The consensus physiological model of the glycoprotein quality control pathway forms a cycle (Fig.~\ref{fig:cycle}), with proteins proceeding through the various states in a directed fashion. This directed protein flux requires free-energy dissipation~\cite{brown2019theory}, representing a cost to cellular resources. To evaluate the impact of this free-energy dissipation on pathway performance, we consider how the folding efficiency depends on the free energy input, for fixed values of protein production rate $k_{\text{pt}}$, misfolded fraction $m_{\text{f}}$, and protein folding speed $k_{\text{f}}$.
The free energy driving the quality control cycle is given by~\cite{brown2019theory}
\begin{equation}
	\label{eq:energy}
	E = k_{\text{B}}T
	\log\frac{k_{\text{c}}k_{\text{r}}k_{\text{g}}}{k_{\text{-c}}k_{\text{-r}}k_{\text{-g}}}	 \ .
\end{equation}
For each value of this
driving energy,
the cycle rate constants are allowed to vary 
so as to maximize the folding efficiency $f_\text{max}^*$.

Figure~\ref{fig:energycycle}a shows that the folding efficiency can increase with the cycle driving energy. In the absence of chaperone-binding background proteins ($P_{\text{b}}=0$), the optimal folding fraction is independent of the energy input into the system. However, when there are background proteins present ($P_{\text{b}}>0$), increasing the energy driving the quality control cycle enables more efficient allocation of chaperone resources specifically to foldable rather than background proteins. For example, reducing the rebinding rate of deglucosylated proteins ($k_{\text{-r}}$) would decrease the fraction of chaperones occupied by background proteins. In the extreme limit $k_{\text{-r}}\rightarrow 0$, background proteins no longer contribute to the system, and the maximal folding efficiency is achieved. However, fully eliminating binding of unglucosylated proteins would require an infinite energy input to provide a fully irreversible process.

In Fig.~\ref{fig:energycycle}b, faster folding (higher $k_{\text{f}}$) leads to a higher folding efficiency, because faster folding can better compete with degradation, and folded proteins free chaperones for other proteins by exiting the cycle.
Both higher protein production ($k_{\text{pt}}$) and misfolded fraction ($m_{\text{f}}$) lead to a lower folding efficiency because fewer chaperones are unoccupied and available for foldable protein binding.

\begin{figure}
	\begin{center}
	\includegraphics[width=0.48\textwidth]{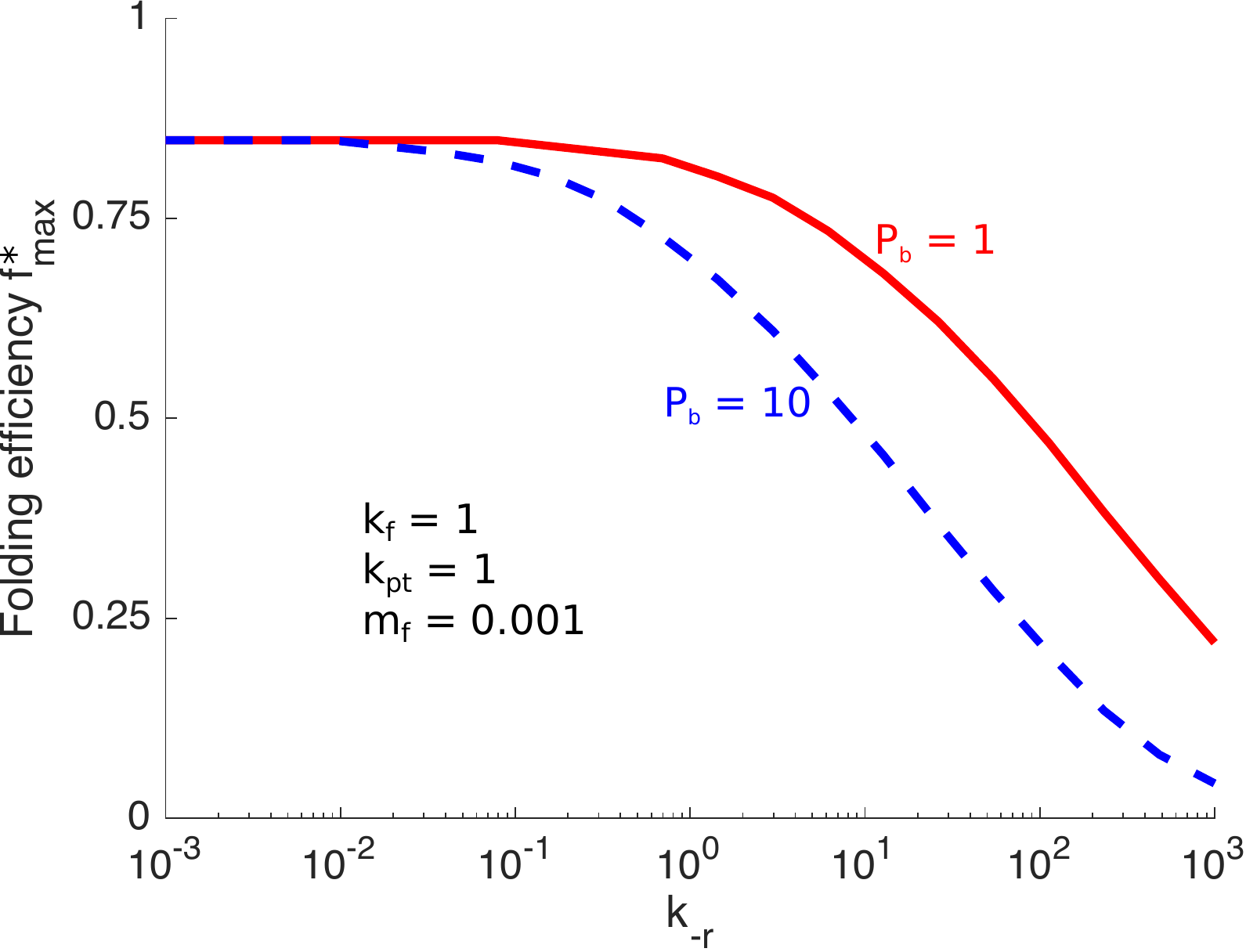}
	\end{center}
	\caption{
		{\bf Untagged protein binding is disadvantageous.}
		Maximal achievable folding fraction at fixed unfolded protein, $P_{\text{tot}}=1$, plotted versus the untagged rebinding rate $k_{\text{-r}}$, as cycle parameters cycle parameters $k_{\text{c}}$, $k_{\text{-c}}$, $k_{\text{r}}$, $k_{\text{g}}$, $k_{\text{-g}}$ are free to vary. Other curves show similar behavior when folding and production rates are altered.
		%
		}
	\label{fig:knr}
\end{figure}

\subsection*{Comparison of performance between models}

A finite driving energy for the quality control cycle implies the presence of reverse processes for all the cycle transitions. We proceed to consider how the presence of the non-physiological reverse transitions for chaperone rebinding $k_{\text{-r}}$ and unbinding $k_{\text{-c}}$ modulates the pathway efficiency.

Figure~\ref{fig:knr} shows that $f_{\text{max}}^*$ monotonically decreases as $k_{\text{-r}}$ increases, for all cases where background proteins are present ($P_{\text{b}}>0$).
This result suggests that removing untagged chaperone binding (i.e.\ setting $k_{\text{-r}}=0$) improves the performance of the chaperone cycle, allowing higher folded protein throughput. Removing untagged binding allows only those proteins recognized as foldable to occupy the chaperone. For the moderate level of background proteins assumed here ($P_\text{b}=1$), this effect becomes small when $k_{\text{-r}}<1$ (corresponding to a rebinding rate smaller than the rate of deglucosylation and chaperone unbinding). However, its importance increases for higher values of $P_{\text{b}}$ (see $P_{\text{b}}=10$ curve in Fig.~\ref{fig:knr}).
Removing the untagged rebinding process entirely can protect the quality control system from potential fluctuations in the total levels of untagged background protein that can result in unproductive chaperone occupation. Having demonstrated the detrimental effects of untagged rebinding, we hereafter set $k_{\text{-r}}=0$, removing this process from the cycle.

\begin{figure*}
	\includegraphics[width=0.96\textwidth]{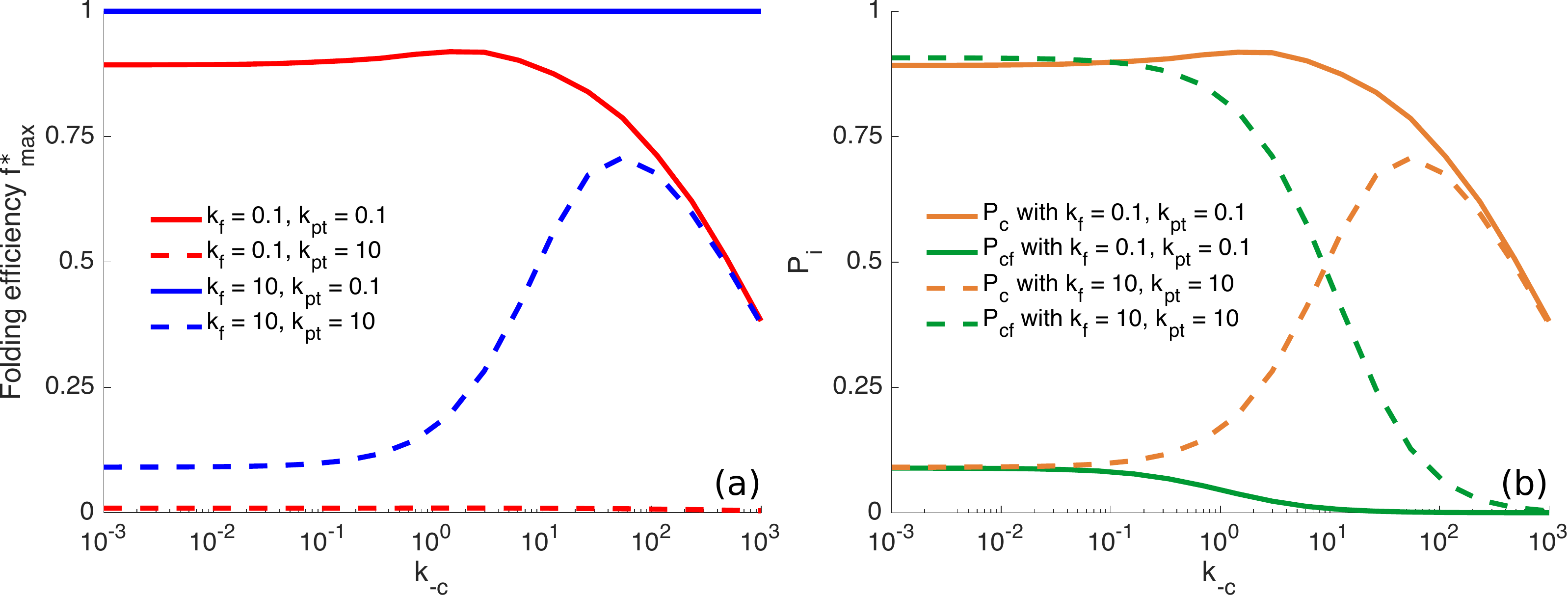}
	\caption{
		{\bf Reversible chaperone binding is disadvantageous at low production rates.}
		(a) Folding efficiency 
		is plotted as a function of unbinding rate $k_{\text{-c}}$.
		Folding rate $k_{\text{f}}$ and production rate $k_{\text{pt}}$ are held constant as indicated.
		(b) Steady-state concentrations of chaperone-bound foldable proteins ($P_{\text{c}}$, orange), and already-folded proteins ($P_{\text{cf}}$, green) for two regimes. Solid lines correspond to low production, slow folding. Dashed lines show high production, rapid folding.
		In both panels, $m_{\text{f}} = 0.001$.
	}
	\label{fig:knc}
\end{figure*}

We next turn our attention to how quality control efficiency varies with $k_{\text{-c}}$, the rate of protein detachment from the chaperone without removal of the glucose tag. For low production and slow folding rates, the folding fraction is maximized or nearly maximized when $k_{\text{-c}}$ is kept low (Fig.~\ref{fig:knc}). In this regime, it is advantageous for the quality control cycle to operate slowly, and high values of $k_\text{-c}\gtrsim 5$ lead to a reduction in the folding fraction by allowing proteins to escape the chaperones before they have a chance to fold. 

By contrast, at high production and fast folding rates, the folding fraction peaks at an intermediate $k_{\text{-c}}$ value. In this regime, rapid turnover through the quality control cycle is advantageous and 
altering the unbinding rate  $k_{\text{-c}}$ leads to two competing effects. On the one hand, more rapid unbinding allows already-folded proteins to be rapidly removed from the chaperones, freeing chaperones to bind other nascent proteins. When the folding process itself is very fast, then already-folded proteins ($P_{\text{cf}}$) can occupy a significant fraction of the available chaperones (Fig.~\ref{fig:knc}b), leading to a decrease in efficiency for low unbinding rates $k_{\text{-c}}$. This effect is not seen for slowly folding proteins, which can be released from the chaperones sufficiently rapidly by the standard deglucosylation pathway ($k_{\text{r}}$).
On the other hand, if the unbinding rate becomes much higher than the folding rate, then there is a tendency for proteins to detach from the chaperone before they can fold, manifesting as low values of $P_{\text{c}}$ (Fig.~\ref{fig:knc}b). Thus, very rapid unbinding 
reduces the efficiency of the system for both rapidly folding and slowly folding cases (Fig.~\ref{fig:knc}a).

Figure~\ref{fig:knc} suggests that different rates of chaperone unbinding ($k_{\text{-c}}$) become optimal in different regimes, depending on whether protein production is sufficiently high and folding is sufficiently slow to overwhelm the available quantity of chaperones.
Chaperones in the ER, such as BiP, are thought to be present in excess quantities~\cite{bakunts2017ratiometric,crofts1998bip,kopp2019upr}, to facilitate rapid chaperone binding of nascent proteins. This suggests that the glycoprotein quality control pathway typically operates in the regime of relatively low production $k_{\text{pt}}\lesssim 1$, so that protein release from chaperones can keep up with the incoming proteins and the chaperones do not become overwhelmed.
At low protein production, Fig.~\ref{fig:knc} shows high values of $k_{\text{-c}}$ primarily decrease the maximum folding fraction. Removing the ability of a protein to detach from a chaperone without glucose trimming should thus improve the performance of the chaperone binding cycle, and folding proteins should be tightly bound to the chaperone until glucose removal. This tight binding may have additional functional importance, such as facilitating recruitment of other enzymes important for folding~\cite{frickel2002trosy,tannous2015n} or as a by-product of the high specificity of chaperone-glucose interaction~\cite{eaton1995let}.

\begin{figure}
	\includegraphics[width=0.96\textwidth]{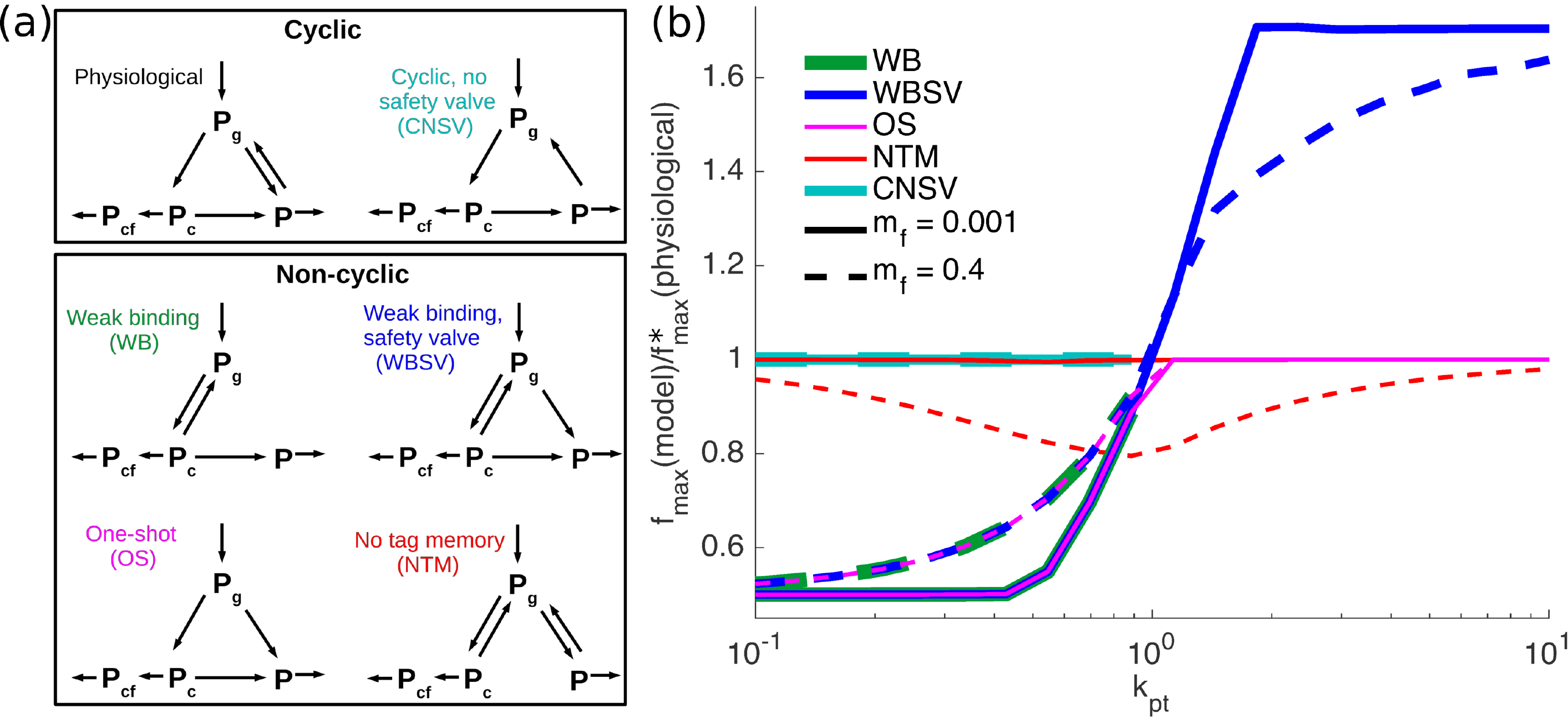}
	\caption{
	{\bf Physiological model for glycoprotein quality control outperforms other models.}
		(a) Schematic of cyclic and non-cyclic models. The physiological model  corresponds to the consensus description of the glycoprotein quality control pathway. 
		(b) Ratios of folding efficiency $f_\text{max}^*$ comparing performance of all models to the physiological model. Curve color indicates the model being compared, with solid lines for $m_{\text{f}}=0.001$ and dashed lines for $m_{\text{f}}=0.4$. For all curves, $k_{\text{f}}=1$.
	}
	\label{fig:models}
\end{figure}

Figures~\ref{fig:knr} and \ref{fig:knc} demonstrate that removing non-specific chaperone binding ($k_{\text{-r}}$) and detachment of proteins from the chaperone without glucose trimming ($k_{\text{-c}}$) improves the performance of the chaperone binding cycle by increasing the maximum folding fraction with a limited accumulation of unfolded protein. The consensus physiological model, with these two processes absent, is thus shown to be more efficient (in the low-production regime) than the full model illustrated in Fig.~\ref{fig:cycle}.

We now explore further glycoprotein quality control pathway model variations, including those that are not cyclic (Fig.~\ref{fig:models}a). The non-cyclic models include all possible variations of a three-state model that lack untagged binding (no $k_{\text{-r}}$) and are capable of producing a finite steady-state solution.
We compare the performance of these models to the consensus physiological model in terms of the efficiency metric $f_\text{max}^*$, at varying levels of protein production (Fig.~\ref{fig:models}b).

The WB (weak binding) model allows proteins to bind and unbind from the chaperone, until the glucose tag is removed.  The WBSV (weak binding, safety valve) model introduces an additional ``safety-valve" pathway where the glucose tag can be removed without chaperone binding. The OS (one shot) model treats chaperone binding and glucose trimming as irreversible, so that each protein only has one chance to attempt folding. These three models share a common feature -- they lack the ability to restore a glucose tag once it is removed, irreversibly committing deglucosylated proteins ($P$) to degradation. Each of these models performs worse than the physiological system in the regime of low protein production ($k_{\text{pt}}<1$), with the folding efficiency  $f_{\text{max}}^*$ dropping by approximately a factor of 2 (Fig.~\ref{fig:models}b). In the regime of high production, the WBSV model is capable of more effectively funneling proteins into a degradation-committed state, allowing it to significantly outperform the physiological model (Fig.~\ref{fig:models}b). However, as discussed previously, cells are believed to typically operate in a regime of limited protein production levels and excess chaperone capacity, so that we focus largely on model performance at low $k_{\text{pt}}$.

In the physiological model, a deglucosylated protein ($P$) is more likely to have first passed through chaperone binding than a monoglucosylated protein ($P_{\text{g}}$). This feature allows glucose moieties to serve as a form of molecular memory -- the presence of a glucose tag means the protein is more likely to be newly made; the absence of the tag means the protein is more likely to have already attempted folding. A contrasting non-cyclic model is the NTM (no tag memory) model, which allows chaperone binding and glucose removal to function as independent processes (Fig.~\ref{fig:models}a). When the fraction of misfolded proteins ($m_\text{f}$) is low, the NTM model performs equivalently to the physiological model. However, when a substantial number of proteins entering the quality control cycle are incapable of being folded (high $m_\text{f}$), the NTM model is at a disadvantage to the physiological system (Fig.~\ref{fig:models}b). In the presence of such defective unfoldable proteins, the cyclic addition and removal of glucose tags allows the physiological model to have a memory of which proteins already attempted (and failed) folding and thus should be made vulnerable to degradation.
Overall, the physiological model outperforms all non-cyclic models in the low-production regime.

The cyclic model with no safety valve (CNSV) exhibits the same cycle as the physiological model: of chaperone binding, deglucosylation upon release, and subsequent reglucosylation (Fig.~\ref{fig:models}a). However, it lacks the direct transition from the tagged state $P_{\text{g}}$ to the vulnerable state $P$.
In the absence of this safety valve pathway, the CNSV model matches the performance of the physiological model at low production rates (Fig.~\ref{fig:models}b). However, for $k_\text{pt}>1$, proteins cannot be released from the chaperones fast enough to keep up with new protein production, and the CNSV model cannot reach a steady-state. In this regime, all chaperones would become clogged with protein and the protein would accumulate indefinitely.
A similar behavior is observed for the non-cyclic WB model, which also lacks the safety-valve (Fig.~\ref{fig:models}b).

\subsection*{Performance and robustness of the physiological model}
\begin{figure*}
	\hspace{-2.2in}
	\includegraphics[width=1.40\textwidth]{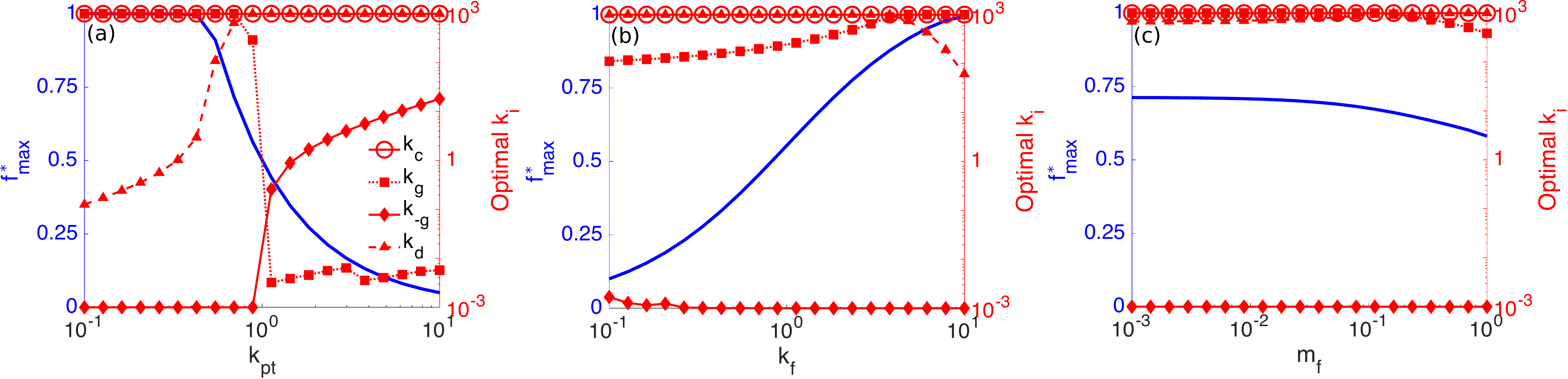}
	\caption{
		{\bf Optimal performance and corresponding parameters.}
		Maximum folding fraction $f_{\text{max}}$ at $P_{\text{tot}} = 1$, (blue curves, left blue vertical axis), and corresponding optimal rate constants, $k_{i}$ (red curves with markers, right red vertical axis) as cycle conditions are varied for the physiological model. (a) varies protein production rate $k_{\text{pt}}$ for fixed $m_{\text{f}} = 10^{-3}$ and $k_{\text{f}} = 1$, (b) varies protein folding rate constant $k_{\text{f}}$ for fixed $m_{\text{f}} = 10^{-3}$ and $k_{\text{pt}} = 0.9$, and (c) varies misfolded fraction $m_{\text{f}}$ for fixed $k_{\text{pt}} = 0.7$ and $k_{\text{f}} = 1$.
	}
	\label{fig:params}
\end{figure*}

We now explore the performance of the physiological model, as well as the optimal kinetic parameter values under different conditions. Performance is quantified in terms of the folding efficiency $f_\text{max}^*$ (the maximum folding fraction at a total protein content $P_{\text{tot}}=1$). We treat the total production rate $k_{\text{pt}}$, protein folding rate $k_\text{f}$, and misfolding fraction $m_\text{f}$ as external input conditions for the system. As always, these rates are expressed relative to the rate of chaperone removal ($k_{\text{r}} = 1$ for non-dimensionalization), which is also treated as fixed.
The quality control pathway is then allowed to adjust all other kinetic rate constants to optimize the folding efficiency -- the resulting optimal folding fraction and the optimized parameters are plotted in Fig.~\ref{fig:params}.

When the overall production rate is low, the optimal folding fraction approaches one (blue curve in Fig.~\ref{fig:params}a), indicating that nearly all the foldable proteins that enter the quality control cycle are successfully folded. At higher production ($k_\text{p}\gtrsim1$), the removal of proteins from chaperones cannot keep up with the flux of incoming proteins. In this regime, the available chaperones in the system are overwhelmed and the folding efficiency drops.

The optimal parameters (red curves in Fig.~\ref{fig:params}a) describe how the optimized quality control system adjusts to changing production rates.
For all conditions explored, binding rate constant ($k_{\text{c}}$) is always maximized, allowing nascent or reglucosylated proteins to bind to chaperones as quickly as possible.
For low $k_{\text{pt}}$, the reglucosylation rate constant $k_{\text{g}}$ is high and the rate constant $k_{\text{-g}}$ for glucose removal from free (not chaperone bound) proteins is low. High $k_{\text{g}}$ and low $k_{\text{-g}}$ indicate that the cycle is quickly removing proteins from the vulnerable state $P$ to prevent degradation, which is expected for low protein production ($k_{\text{pt}}$) and low misfolded fraction ($m_{\text{f}}$) as proteins will then usually be provided multiple rounds of chaperone binding. In this regime, the optimal degradation rate ($k_{\text{d}}$) rises gradually with increasing production in order to maintain a constant amount of unfolded protein $P_{\text{tot}}=1$.
Eventually (when $k_\text{pt}\rightarrow 1$) there will not be sufficient chaperones to fold all proteins, and protein degradation must increase sharply to maintain a fixed level of total unfolded protein. 

As the production rate passes $k_{\text{pt}} \approx 1$,
protein reglucusylation ($k_{\text{g}}$) steeply decreases and glucose removal ($k_{\text{-g}}$) increases. This switch indicates the activation of
the `safety valve' pathway which moves excess proteins directly into the degradation-vulnerable state $P$ to avoid accumulation of unfolded proteins.
As protein production continues to increase, glucose removal via $k_{\text{-g}}$ further increases to enhance this safety valve. Overall, there are two regimes: low protein production, where chaperones are available and proteins are quickly tagged for chaperone rebinding to prioritize folding; and high protein production, where chaperones are overwhelmed and rapid deglucosylation and degradation is prioritized.

Figure~\ref{fig:params}b shows how performance and optimal parameters change as protein folding speed $k_{\text{f}}$ is varied. As expected, the folding fraction increases with folding speed. The increased folding speed does not cause significant changes in the optimal parameters, with a modest increase in reglucosylation ($k_{\text{g}}$) and decreases in glucose removal ($k_{\text{-g}}$) and degradation ($k_{\text{d}}$) as faster folding frees up chaperones. Figure~\ref{fig:params}c shows that increasing the misfolded fraction $m_{\text{f}}$ modestly decreases the folding efficiency while leaving optimal parameters largely unchanged.

\begin{figure}
	\includegraphics[width=0.96\textwidth]{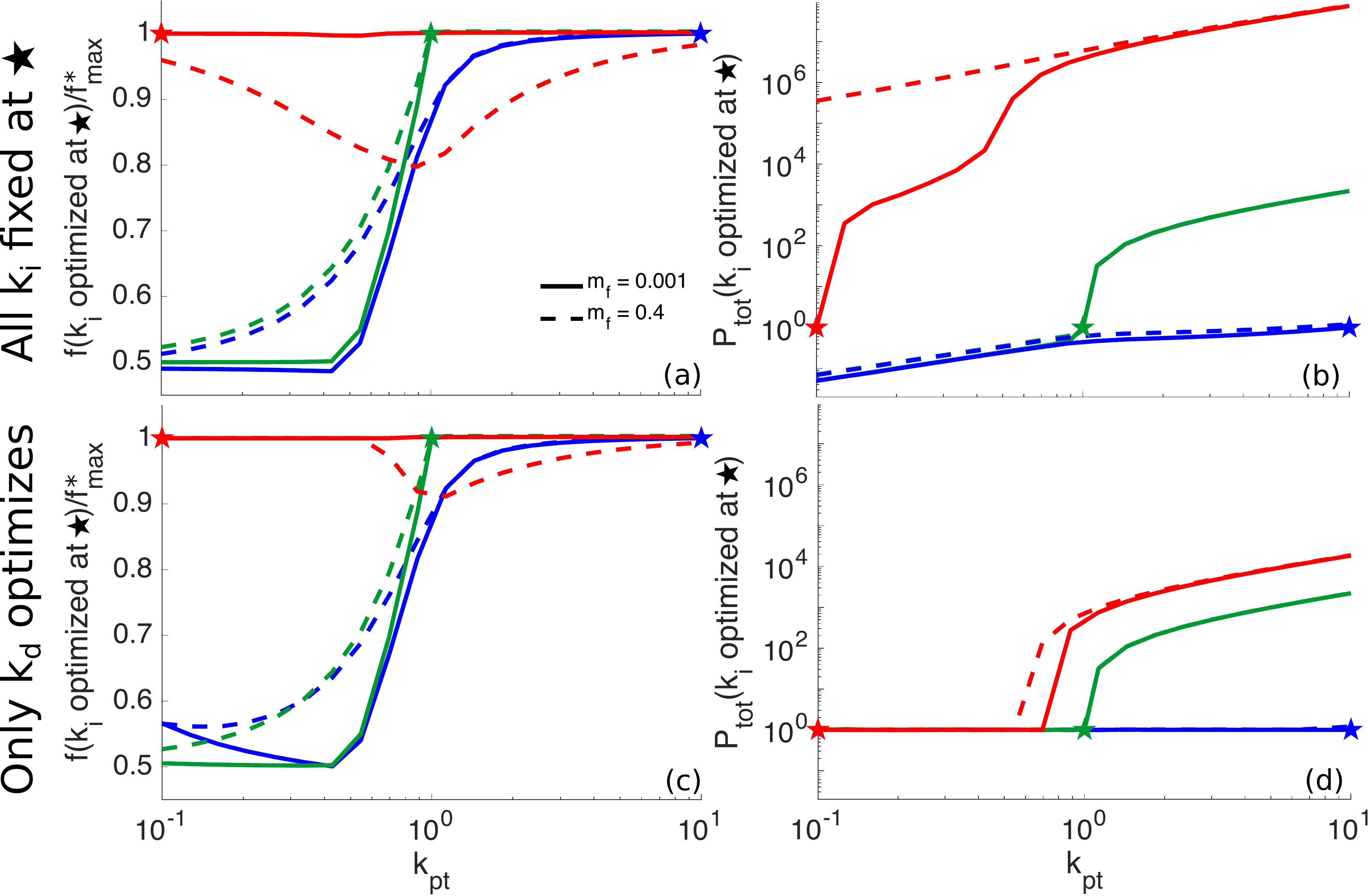}	
	\caption{
		{\bf Robustness of quality control cycle to changing production rates.}
		(a) The folding fraction achieved with fixed rate constants is plotted as a fraction of the maximum achievable folding fraction $f_{\text{max}}$ for $P_{\text{tot}}=1$ as the protein production rate $k_{\text{pt}}$ is varied. Fixed rate constants $k_{\text{c}}$, $k_{\text{g}}$, $k_{\text{-g}}$, and $k_{\text{d}}$ are those that achieve $f_{\text{max}}$ with $m_{\text{f}}=0.001$ at various protein production levels: $k_{\text{pt}}=0.1$ (red curves and star), $k_{\text{pt}}=1$ (green), and $k_{\text{pt}}=10$ (blue). (b) The $P_{\text{tot}}$ corresponding to the folding fractions in (a). (c) Analogous plot to (a), with rate constants fixed at the optimal values for specific $k_{\text{pt}}$ values, except that the degradation rate constant $k_{\text{d}}$ adjusts to maintain $P_{\text{tot}}=1$. If $k_{\text{d}}$ adjustment cannot achieve $P_{\text{tot}}=1$, then $k_{\text{d}}$ adjustment minimizes the difference from $P_{\text{tot}}=1$. (d) The $P_{\text{tot}}$ corresponding to the folding fractions in (c).
	}
	\label{fig:sensitivity}
\end{figure}

Overall, Fig.~\ref{fig:params} demonstrates that maintaining maximum folding efficiency requires large variation in parameters if the protein production level changes, but limited variation in parameters for changes to protein folding speed and misfolded protein fraction. We next proceed to explore how well the cycle can perform under changing production levels if a single fixed parameter set is used across all values of $k_{\text{pt}}$.
The goal is to assess the robustness of this quality control system to changes in protein production, for the case where other parameters cannot be adjusted sufficiently rapidly to keep up with such changes.

We consider the robustness of a fixed quality control system as follows.
The rate constants are optimized to give maximal folding efficiency for a given value of input conditions $k_\text{pt}$, $k_\text{f}$, $m_\text{f}$. For those parameters and input conditions, the system gives the highest folding fraction $f_{\text{max}}^*$ that maintains a fixed protein content $P_\text{tot}=1$. When the input production rate $k_{\text{pt}}$ is varied
and all remaining parameters are held fixed, the folding fraction will decrease below this optimum value (Fig.~\ref{fig:sensitivity}a) and the total protein content $P_\text{tot}$ will also change (Fig.~\ref{fig:sensitivity}b). The values plotted in Fig.~\ref{fig:sensitivity}a,b are given relative to the folding fraction and protein content at the point where the system was optimized.

If the parameters are optimized at low protein production ($k_{\text{pt}}=0.1$), the system continues to achieve close to the optimal folding fraction when the protein production is  increased (Fig.~\ref{fig:sensitivity}a, red curves). However,  the total accumulated protein increases by orders of magnitude even for a modest rise in the production rate (Fig.~\ref{fig:sensitivity}b, red curves). 
If the parameters are optimized at high  protein production ($k_{\text{pt}}=10$), and the production rate is lowered significantly, then the folding efficiency is reduced to roughly half of the optimal amount and the accumulated protein is also decreased (Fig.~\ref{fig:sensitivity}a,b, blue curves). A system optimized at intermediate production ($k_{\text{pt}}=1$) exhibits analogous behaviors (Fig.~\ref{fig:sensitivity}a,b,  green curves). If the production rate is lowered, the folded fraction drops below optimal values. If
raised, then a massive increase in accumulated protein is observed. 

These results highlight a general principle: the quality control cycle can be optimized to operate in one of two regimes: a regime with excess chaperone capacity, and one where the chaperones are overwhelmed. Optimizing for the former requires shutting off the safety-valve, and prioritizing reglucosylation over degradation. Optimizing for the latter requires enhancing degradation and deglucosylation. The transition between the two regimes occurs when the rate of protein production becomes comparable to the rate at which chaperone-bound proteins detach from chaperones (i.e.: at $k_\text{pt} = 1$). A system optimized for low production will result in large-scale protein accumulation if the production rate is increased by even a modest amount. A system optimized for high production yields suboptimal folding throughput if shifted to the low-production regime. 

Without any flexibility to adjust cycle parameters, the glycoprotein quality control system will perform poorly in one of the two regimes. A natural question is to what extent adjusting a single kinetic parameter will allow the system to compensate for changing production rates and to perform well across a broad range of conditions.
Figure~\ref{fig:params}a shows that the optimal degradation rate ($k_{\text{d}}$) continuously changes across a range of low protein production levels ($k_{\text{pt}}$), suggesting $k_{\text{d}}$ as a good candidate for an adjustable parameter.
Thus we choose to treat the degradation rate $k_{\text{d}}$ as capable of adapting to changing production levels, while all other rate constants in the cycle are held fixed. At each value of the production rate, $k_{\text{d}}$ is adjusted to maintain a total protein content $P_\text{tot}=1$ whenever possible, with the resulting folding fraction shown in Fig.~\ref{fig:sensitivity}c. 
For a system optimized at low protein production, an adjustable degradation rate allows the optimum folding fraction to be maintained across all production rates. Even when the fraction of misfolded proteins is increased (dashed curves in Fig.~\ref{fig:sensitivity}c), the optimum folding fraction can be maintained up to intermediate production levels. 

Strikingly, a system optimized at low protein production can also maintain a fixed total protein $P_{\text{tot}}=1$ up to intermediate production levels ($k_\text{p}\lesssim 0.7$) by adjusting the degradation rate $k_{\text{d}}$ (Fig.~\ref{fig:sensitivity}d).
The ability of this system to maintain fixed total protein content over a broad range of low to intermediate production values is in sharp contrast to the rapidly increasing protein levels that arise when all parameters are held fixed (Fig.~\ref{fig:sensitivity}b). 
At higher production rates, there is no value of the degradation rate that can maintain the fixed total protein content and we adjust $k_{\text{d}}$ as needed to minimize $P_\text{tot}$.
Allowing $k_{\text{d}}$ to adjust in a system optimized for intermediate or high protein production  has little impact on both folding fraction and protein accumulation compared to the fully fixed system (Fig.~\ref{fig:sensitivity}c,d).

This analysis establishes that the quality control pathway can perform well at typical low production rates, yet be capable of adapting to moderate surges in protein production. Such robust behavior requires only for the protein degradation rate to be rapidly adjustable to changing conditions. 
Other parameters in the quality control cycle can be held constant while allowing near-optimal system performance over an order of magnitude range in protein production. Interestingly,
there is evidence that cellular quality control systems do in fact control protein degradation throughput in response to perturbations in protein homeostasis. Specifically, cells maintain a reservoir of ERAD enzymes 
in ER-associated vesicles that can fuse with the ER lumen in response to an accumulation of unfolded proteins, rapidly upregulating protein degradation~\cite{benyair2015glycan,benyair2015mammalian,ogen2016mannosidase}. 

\section*{Discussion}
We have investigated the impact of pathway architecture and kinetic parameters on the performance of the glycoprotein quality control cycle in the endoplasmic reticulum.
Two metrics are used to evaluate steady-state performance. The fraction of foldable proteins that are successfully folded measures the accuracy of the system. The total quantity of unfolded proteins measures processing speed, with lower protein levels corresponding to more rapid processing.

Broadly, we find that a cyclic quality control process, with protein substrates driven in a preferred direction through three quality control states, leads to improved performance. Energy is required for cyclic driving, and increased driving energy per cycle allows higher protein folding fractions (Fig.~\ref{fig:energycycle}). A higher folding fraction is achieved by eliminating reverse transitions that are absent from the consensus physiological model of the glycoprotein quality control pathway (Figs.~\ref{fig:knr} and \ref{fig:knc}). This matches the directed, cyclic behavior commonly described as occurring for physiological glycoprotein quality control.

The energy-consuming nature of the quality control cycle improves its decision making during the protein folding process, echoing other examples of biomolecular processes cyclically driven out of equilibrium to improve their performance. DNA copying is famously driven out of equilibrium in a `kinetic proofreading' process that increases its accuracy~\cite{hopfield1974kinetic,ninio1975kinetic}. Similar cyclic nonequilibrium processes increase the accuracy of T-cell signaling~\cite{mckeithan1995kinetic} and sensing of external concentrations~\cite{mehta2012energetic}.

By exhaustively considering all remaining cyclic and non-cyclic variations of the glycoprotein quality control pathway, we show that the consensus physiological model outperforms all other viable models (Fig.~\ref{fig:models}). Models lacking a `safety valve', or a path for protein degradation without chaperone binding, will dangerously accumulate unfolded proteins at high protein production levels. This safety valve requirement aligns with the only two-way transition in the consensus physiological model, which allows glucose tags to be removed from proteins that are not bound the chaperone, facilitating their degradation.

We find that the optimal tuning
of the consensus physiological model varies substantially with protein production level (Fig.~\ref{fig:params}a). If the cell must choose a particular set of rate constants, it will either sacrifice folded protein throughput at low protein production levels, or induce massive unfolded protein accumulation at higher production levels (Fig.~\ref{fig:sensitivity}a,b). 
A particularly robust system design requires optimizing parameters for low protein production and allowing a single rate constant (the degradation rate) to adapt to changing production levels. Such a system can successfully maintain both maximum folding efficiency and low unfolded protein accumulation across a range of low-to-intermediate production rates (Fig.~\ref{fig:sensitivity}c,d).

\textit{In vivo} glycoprotein folding in the ER is thought to operate in a low protein production regime, matching the robust system design. Namely, there is excess protein folding capacity in the ER under basal conditions~\cite{bakunts2017ratiometric}, with abundant chaperones that exceed the requirements of the protein folding load~\cite{crofts1998bip,kopp2019upr}. Under conditions of  high protein folding load, chaperones are overwhelmed~\cite{bergmann2018three,oikonomou2020disposing}, and the unfolded protein response is triggered, driving down the effective protein folding load by increasing chaperone quantity~\cite{bakunts2017ratiometric} and reducing protein translation~\cite{preissler2019early}.

The adjustable degradation rate, which alone can maintain both high folding throughput and low unfolded protein accumulation, corresponds to the dynamic behavior observed for some ERAD enzymes that remove proteins from the ER for degradation.
Certain mannosidases, important for ERAD targeting, are largely sequestered to quality control vesicles in the absence of ER stress~\cite{benyair2015glycan,benyair2015mammalian,ogen2016mannosidase}. When the ER becomes stressed (i.e.\ unfolded proteins accumulate), these mannosidases converge on the ER, rapidly increasing degradation targeting~\cite{benyair2015glycan,benyair2015mammalian,ogen2016mannosidase}.
Comparison of timescales for mannosidase convergence to the ER following proteasome inhibition (approximately a couple hours~\cite{bakunts2017ratiometric}) and the gene expression response to accumulation of unfolded proteins (approximately 5 to 10 hours~\cite{benyair2015mammalian}) indeed suggests that ERAD-mediated degradation may be enhanced relatively quickly.

In contrast to the large variations in optimal rate constants with protein production level, changes in protein folding speed require relatively little variation to the optimal rate constants (Fig.~\ref{fig:params}b). The glycoprotein quality control pathway must simultaneously process a variety of proteins, which can have folding times ranging from a few minutes to several hours~\cite{hebert2007and}. The ability of a single pathway to near-optimally process this variety of folding speeds appears to be a strength of its design. The efficiency of protein throughput can approach 100$\%$ and range down to 25$\%$ or lower for slow-folding proteins or proteins with mutations~\cite{hebert2007and}.
Fig.~\ref{fig:params}b suggests that these low efficiencies (ranging down to 25$\%$ or lower) are not the result of a poorly-tuned quality control process, but instead that the low efficiencies are an unavoidable consequence of slow folding.

The optimal rate constants also change little with the fraction of produced proteins which are inherently misfolded or unfoldable (Fig.~\ref{fig:params}c). This suggests the design of the quality control pathway is robust to the onset of systematic misfolding, which may arise from  translation errors, environmental stress, or mutations~\cite{tyedmers2010cellular}, so long as the total protein production levels remain relatively unchanged.

Effective quality control of glycoprotein folding in the endoplasmic reticulum ensures an adequate supply of functional natively-folded proteins and limits the accumulation of misfolded proteins. The failure to provide sufficient natively-folded proteins~\cite{hebert2007and} and the formation of misfolded protein aggregates~\cite{soto2003unfolding} can both contribute to the onset of disease. Our modeling quantitatively demonstrates how the performance of this pathway under a broad range of conditions is modulated by key kinetic parameters that serve as potential targets for pharmacological or genetic perturbations.
This quantitative framework serves as a basic foundation for understanding the glycoprotein quality control pathway, which can be further expanded in future work to account for more complex aspects, such as sequential glycan sugar moiety removal~\cite{benyair2015mammalian,shenkman2018mannosidase} and the spatial organization of quality control activities~\cite{benyair2015glycan}.

\subsection*{Methods}

\subsection*{Non-dimensionalization}

We non-dimensionalize all times by the timescale of protein removal from the chaperone via glucose trimming, $k_{\text{r}}^{-1}$, and all concentrations by total chaperone concentration, $C_{\text{tot}}$. For conciseness of notation, all kinetic parameters in the text refer to non-dimensionalized values. The dimensionless dynamic equations for foldable proteins are then:
\begin{subequations}
	\label{eq:desapp}
	\begin{align}
		\frac{\text{d}P_{\text{g}}}{\text{d}t} &= k_{\text{g}}P + k_{\text{-c}}P_{\text{c}} - (k_{\text{c}}C_{\text{A}} + k_{\text{-g}})P_{\text{g}} + k_{\text{p}} \\
		\frac{\text{d}P_{\text{c}}}{\text{d}t} &= (k_{\text{c}}P_{\text{g}} + k_{\text{-r}}P)C_{\text{A}} - (1 + k_{\text{-c}} + k_{\text{f}})P_{\text{c}} \\
		\frac{\text{d}P}{\text{d}t} &= P_{\text{c}} + k_{\text{-g}}P_{\text{g}} - (k_{\text{g}} + k_{\text{-r}}C_{\text{A}} - k_{\text{d}})P \\
		\frac{\text{d}P_{\text{cf}}}{\text{d}t} &= k_{\text{f}}P_{\text{c}} - (1 + k_{\text{-c}})P_{\text{cf}} \ , \\
		\frac{\text{d}P_{\text{cb}}}{\text{d}t} &= k_{\text{-r}}C_{\text{A}}P_{\text{b}} - k_{\text{r}}P_{\text{cb}} \ .
	\end{align}
\end{subequations}
The dynamics of misfolded proteins are described by
\begin{subequations}
	\label{eq:desapp2}
	\begin{align}
		\frac{\text{d}P_{\text{g}}^*}{\text{d}t} &= k_{\text{g}}P^* + k_{\text{-c}}P_{\text{c}}^* - (k_{\text{c}}C_{\text{A}} + k_{\text{-g}})P_{\text{g}}^* + k_{\text{p}}^* \\
		\frac{\text{d}P_{\text{c}}^*}{\text{d}t} &= (k_{\text{c}}P_{\text{g}}^* + k_{\text{-r}}P^*)C_{\text{A}} - (1 + k_{\text{-c}})P_{\text{c}}^* \\
		\frac{\text{d}P^*}{\text{d}t} &= P_{\text{c}}^* + k_{\text{-g}}P_{\text{g}}^* - (k_{\text{g}} + k_{\text{-r}}C_{\text{A}} - k_{\text{d}})P^* \ .
	\end{align}
\end{subequations}
Note that most rate constants are the same for both foldable and misfolded proteins, except $k_{\text{p}}$ changes to $k_{\text{p}}^*$ to allow different production rates of foldable and misfolded proteins, $k_{\text{f}}^*=0$ (misfolded proteins cannot fold), and $P_{\text{b}}^*=0$ (as only a single comprehensive population of background proteins is considered).
The available amount of chaperone is $C_{\text{A}} = 1- P_{\text{c}} - P_{\text{cf}} - P_{\text{c}}^* - P_{\text{cb}}$, where $C_{\text{tot}}=1$ is the dimensionless total chaperone concentration.

\subsection*{Steady-state solution of chaperone cycle dynamics}

Equations~\ref{eq:desapp} and \ref{eq:desapp2}
describe the dynamics of the chaperone folding cycle. In steady state each of the time derivatives must equal zero. By summing together Eqs.~\ref{eq:desapp}a--d, we get the steady-state condition for the total flux of foldable proteins through the cycle:
\begin{equation}
	\label{eq:ssinout1}
	k_{\text{p}} = k_{\text{f}}P_{\text{c}} + k_{\text{d}}P \ .
\end{equation}
Rearranged, this gives $P$ in terms of parameters and $P_{\text{c}}$,
\begin{equation}
	\label{eq:ssinout}
	P = \frac{k_{\text{p}}}{k_{\text{d}}} - \frac{k_{\text{f}}}{k_{\text{d}}}P_{\text{c}} \ .
\end{equation}
Applying $\text{d}P_{\text{cf}}/\text{d}t = 0$ gives
\begin{equation}
	\label{eq:sscf}
	P_{\text{cf}} = \frac{k_{\text{f}}}{k_{\text{r}}+k_{\text{-c}}}P_{\text{c}} \ ,
\end{equation}
and $\text{d}P_{\text{cb}}/\text{d}t = 0$ gives
\begin{equation}
	\label{eq:sscb}
	P_{\text{cb}} = k_{\text{-r}}P_{\text{b}}C_{\text{A}} \ ,
\end{equation}
where the available chaperone is
\begin{subequations}
	\begin{align}
		C_{\text{A}} &\equiv 1-P_{\text{c}}-P_{\text{cf}}-P_{\text{c}}^* - P_{\text{cb}} \\
		&= 1 - \left(1 + \frac{k_{\text{f}}}{k_{\text{r}} + k_{\text{-c}}}\right)P_{\text{c}} - P_{\text{c}}^* - k_{\text{-r}}P_{\text{b}}C_{\text{A}} \ . \label{eq:availC2}
	\end{align}
\end{subequations}

Substituting Eqs.~\ref{eq:ssinout}, \ref{eq:sscf}, and \ref{eq:sscb} into Eqs.~\ref{eq:desapp}a,b gives
\begin{subequations}
	\label{eq:dealpha}
	\begin{align}
		\frac{\text{d}P_{\text{c}}}{\text{d}t} &= C_{\text{A}}\left(k_{\text{c}}P_{\text{g}} + \frac{k_{\text{-r}}k_{\text{p}}}{k_{\text{d}}} - \frac{k_{\text{-r}}k_{\text{f}}}{k_{\text{d}}}P_{\text{c}}\right) - (k_{\text{f}}+1+k_{\text{-c}})P_{\text{c}}\\
		\frac{\text{d}P_{\text{g}}}{\text{d}t} &= k_{\text{p}} + k_{\text{-c}}P_{\text{c}} + \frac{k_{\text{g}}k_{\text{p}}}{k_{\text{d}}} - \frac{k_{\text{g}}k_{\text{f}}}{k_{\text{d}}}P_{\text{c}} -k_{\text{-g}}P_{\text{g}} - C_{\text{A}} k_{\text{c}}P_{\text{g}} \ .
	\end{align}
\end{subequations}

Similarly for misfolded proteins, which have $k_{\text{f}}^*=0$, the steady state condition for protein fluxes entering and exiting the cycle is
\begin{equation}
	\label{eq:ssinout2}
	P^* = \frac{k_{\text{p}}^*}{k_{\text{d}}} \ .
\end{equation}
Substituting Eq.~\ref{eq:ssinout2} into Eqs.~\ref{eq:desapp2}a,b gives
\begin{subequations}
	\label{eq:dealpha2}
	\begin{align}
		\frac{\text{d}P_{\text{c}}^*}{\text{d}t} &= C_{\text{A}}\left(k_{\text{c}}P_{\text{g}}^* + \frac{k_{\text{-r}}k_{\text{p}}^*}{k_{\text{d}}}\right) - (1+k_{\text{-c}})P_{\text{c}}^*\\
		\frac{\text{d}P_{\text{g}}^*}{\text{d}t} &= k_{\text{p}}^* + k_{\text{-c}}P_{\text{c}}^* + \frac{k_{\text{g}}k_{\text{p}}^*}{k_{\text{d}}} -k_{\text{-g}}P_{\text{g}}^* - C_{\text{A}} k_{\text{c}}P_{\text{g}}^* \ .
	\end{align}
\end{subequations}

Equation~\ref{eq:dealpha} can be rewritten as $\mathbf{M}\vec{P} = \vec{b}$:
\begin{equation}
	\mathbf{M}\vec{P} =
	\begin{bmatrix}
		C_{\text{A}} k_{\text{c}} & C_{\text{A}} m_1+n_1 \\
		-C_{\text{A}} k_{\text{c}} - k_{\text{-g}} & n_2
	\end{bmatrix}
	\begin{bmatrix}
		P_{\text{g}}\\
		P_{\text{c}}
	\end{bmatrix}
	=
	\begin{bmatrix}
		C_{\text{A}} b_1\\
		b_2
	\end{bmatrix}
	=\vec{b} \ ,
\end{equation}
with $m_1 = -k_{\text{-r}}k_{\text{f}}/k_{\text{d}}$, $n_1 = -(k_{\text{f}} + 1 + k_{\text{-c}})$, $n_2 = k_{\text{-c}} - k_{\text{g}}k_{\text{f}}/k_{\text{d}}$, $b_1 = -k_{\text{-r}}k_{\text{p}}/k_{\text{d}}$, and $b_2 = -(k_{\text{p}} + k_{\text{g}}k_{\text{p}}/k_{\text{d}})$. The determinant $|\mathbf{M}|=k_{\text{c}}m_1C_{\text{A}}^2 + (k_{\text{-g}}m_1 + k_{\text{c}}n_1 + k_{\text{c}}n_2)C_{\text{A}} + k_{\text{-g}}n_1 = r_2C_{\text{A}}^2 + r_1C_{\text{A}} + r_0$.
Rearranging gives
\begin{equation}
	\label{eq:pc1}
	\begin{bmatrix}
		P_{\text{g}}\\
		P_{\text{c}}
	\end{bmatrix}
	=
	\frac{1}{r_2C_{\text{A}}^2 + r_1C_{\text{A}} + r_0}
	\begin{bmatrix}
		p_1C_{\text{A}} + p_0 \\
		q_2C_{\text{A}}^2 + q_1C_{\text{A}}
	\end{bmatrix} \ ,
\end{equation}
with $p_1 = b_1n_2-m_1b_2$, $p_0 = -b_2n_1$, $q_2 = k_{\text{c}}b_1$, $q_1 = k_{\text{-g}}b_1 + k_{\text{c}}b_2$.

Similarly, Eq.~\ref{eq:dealpha2} can be rewritten as $\mathbf{M^*}\vec{P}^*=\vec{b}^*$:
\begin{equation}
	\mathbf{M^*}\vec{P^*} =
	\begin{bmatrix}
		C_{\text{A}} k_{\text{c}} & n_1^* \\
		-C_{\text{A}} k_{\text{c}} - k_{\text{-g}} & k_{\text{-c}}
	\end{bmatrix}
	\begin{bmatrix}
		P_{\text{g}}^*\\
		P_{\text{c}}^*
	\end{bmatrix}
	=
	\begin{bmatrix}
		C_{\text{A}} b_1^*\\
		b_2^*
	\end{bmatrix}
	=\vec{b} \ ,
\end{equation}
with $n_1^* = -(1 + k_{\text{-c}})$, $b_1^* = -k_{\text{-r}}k_{\text{p}}^*/k_{\text{d}}$, and $b_2^* = -k_{\text{p}}^* - k_{\text{g}}k_{\text{p}}^*/k_{\text{d}}$. The determinant $|\mathbf{M}^*| = C_{\text{A}}(k_{\text{c}}k_{\text{-c}} + n_1^*k_{\text{c}}) + n_1^*k_{\text{-g}} = C_{\text{A}} r_1^* + r_0^*$. Rearranging gives
\begin{equation}
	\label{eq:pc2}
	\begin{bmatrix}
		P_{\text{g}}^*\\
		P_{\text{c}}^*
	\end{bmatrix}
	= 
	\frac{1}{C_{\text{A}} r_1^* + r_0^*}
	\begin{bmatrix}
		p_1^*C_{\text{A}} + p_0^*\\
		q_2^*C_{\text{A}}^2 + q_1^*C_{\text{A}}
	\end{bmatrix} \ ,
\end{equation}
with $p_1^* = k_{\text{-c}}b_1^*$, $p_0^* = -b_2^*n_1^*$, $q_2^* = k_{\text{c}}b_1^*$, and $q_1^* = k_{\text{-g}}b_1^* + k_{\text{c}}b_2^*$.

We now insert $P_{\text{c}}$ from Eq.~\ref{eq:pc1} and $P_{\text{c}}^*$ from Eq.~\ref{eq:pc2} into Eq.~\ref{eq:availC2},
\begin{align}
	C_{\text{A}} &= 1 - \left(1 + \frac{k_{\text{f}}}{1 + k_{\text{-c}}}\right)\frac{q_2C_{\text{A}}^2 + q_1C_{\text{A}}}{r_2C_{\text{A}}^2 + r_1C_{\text{A}} + r_0} \nonumber \\
	&- \frac{q_2^*C_{\text{A}}^2 + q_1^*C_{\text{A}}}{r_1^*C_{\text{A}} + r_0^*} - \frac{k_{\text{-r}}}{k_{\text{r}}}P_{\text{b}}C_{\text{A}} \ .
\end{align}
Rearranging,
\begin{align}
	\label{eq:solutionfinal}
	&C_{\text{A}}(1 + k_{\text{-r}}P_{\text{b}})[r_2r_1^*C_{\text{A}}^3 + (r_2r_0^* + r_1r_1^*)C_{\text{A}}^2 \nonumber \\
	+ &(r_1r_0^* + r_0r_1^*)C_{\text{A}} + r_0r_0^*] \nonumber \\
	- &[r_2r_1^*C_{\text{A}}^3 + (r_2r_0^* + r_1r_1^*)C_{\text{A}}^2 + (r_1r_0^* + r_0r_1^*)C_{\text{A}} + r_0r_0^*] \nonumber \\
	+ &[1 + k_{\text{f}}/(k_{\text{-c}}+1)](q_2C_{\text{A}}^2 + q_1C_{\text{A}})(C_{\text{A}} r_1^* + r_0^*) \nonumber \\
	+ &(q_2^*C_{\text{A}}^2 + q_1^*C_{\text{A}})(r_2C_{\text{A}}^2 + r_1C_{\text{A}} + r_0) = 0 \ .
\end{align}
This forms a quartic equation for $C_{\text{A}}$, which can be solved with standard root-finding algorithms.
Once $C_{\text{A}}$ is obtained, Eqs.~\ref{eq:pc1} and \ref{eq:pc2} give steady state $P_{\text{c}}$, $P_{\text{g}}$, $P_{\text{c}}^*$, and $P_{\text{g}}^*$, from which Eq.~\ref{eq:ssinout} gives steady state $P$. Eq.~\ref{eq:ssinout2} gives steady state $P^*$ once $k_{\text{p}}^*$ and $k_{\text{d}}$ are selected, without needing other information.

\subsection*{Optimization of cycle efficiency}

For the results in Fig.~\ref{fig:phasediagram}, the maximum folding fraction independent of total unfolded protein was first found by allowing $k_i = k_{\text{c}}$, $k_{\text{-c}}$, $k_{\text{g}}$, $k_{\text{-g}}$, $k_{\text{-r}}$, and $k_{\text{d}}$ to vary to maximize the folding fraction using the Matlab routine fmincon, with $k_i\in[10^{-3},10^3]$. The minimum total unfolded protein is found using the Matlab routine fmincon for each fixed folding fraction (at a value less than or equal to the maximum folding fraction), constrained with the nonlinear constraints option, and with $k_i\in[10^{-3},10^3]$.

For the results in Fig.~\ref{fig:energycycle}, $k_i = k_{\text{c}}$, $k_{\text{-c}}$, $k_{\text{-r}}$, $k_{\text{g}}$, $k_{\text{-g}}$, and $k_{\text{d}}$ are allowed to vary to maximize the folding fraction (Eq.~\ref{eq:foldfrac}), while fixing the energy (Eq.~\ref{eq:energy}) at a specific value, and fixing the total unfolded protein $P_{\text{tot}}=1$. The folding fraction maximization was performed using the Matlab routine fmincon with energy and total unfolded protein fixed using the nonlinear constraints option. The $k_{i}$ were free within the range $k_i\in[10^{-3},10^3]$.

The results in Fig.~\ref{fig:knr} are found similarly to those of Fig.~\ref{fig:phasediagram}, with the fixed folding fraction varied using the bisection method until a $P_{\text{tot}}\in(0.99,1.01)$ is found. Results in Figs.~\ref{fig:knc} and \ref{fig:params} are found with the same method as Fig.~\ref{fig:knr} with the appropriate $k_i$ set to zero and the appropriate $k_{i}$ allowed to vary within $k_i\in[10^{-3},10^3]$. Almost all results in Fig.~\ref{fig:models} are also found with the method of Figs.~\ref{fig:knc} and \ref{fig:params}. The exception in Fig.~\ref{fig:models} is the no tag memory (NTM) model, which lacks the transition represented with rate constant $k_{\text{r}}$, and instead sets $k_{\text{-c}}=1$.

The $f_{\text{max}}^*$ and optimizing $k_i^*$ at particular $k_{\text{pt}}$ in Fig.~\ref{fig:sensitivity} are found with the same method as Figs.~\ref{fig:knr}, \ref{fig:knc}, \ref{fig:models}, and \ref{fig:params}. The optimal parameters that achieve $f_{\text{max}}^*$ are then used as the fixed parameters in Eq.~\ref{eq:solutionfinal} to determine the folding fraction and total unfolded protein in Fig.~\ref{fig:sensitivity}a,b as the protein production is varied. The folding fraction and total unfolded protein in Fig.~\ref{fig:sensitivity}c,d with only $k_{\text{d}}$ free is found by using the bisection method to vary $k_{\text{d}}$ to attempt to find a $k_{\text{d}}$ value with $P_{\text{tot}}=1$. If $P_{\text{tot}}=1$ cannot be achieved with $k_{\text{d}}\in[10^{-3},10^3]$ then $k_{\text{d}}=10^3$ is chosen to minimize $P_{\text{tot}}$. 

\section*{Acknowledgments}
This work was supported in part by funding from the Hellman Fellows Fund, the Alfred P. Sloan Foundation, and a Cottrell Scholars Award from the Research Corporation for Science Advancement.

\nolinenumbers

%
%
%


\end{document}